\documentclass[preprint,superscriptaddress,showpacs,preprintnumbers,amsmath,amssymb]{revtex4}
\usepackage{booktabs}
\usepackage{graphicx,color}% Include figure files
\usepackage{dcolumn}% Align table columns on decimal point
\usepackage{bm}% bold math
\usepackage{CJK}
\usepackage{mathrsfs}
\usepackage[dvipdfm,
            pdfstartview=FitH,
            CJKbookmarks=true,
            bookmarksnumbered=true,
            bookmarksopen=true,
            colorlinks,
            pdfborder=001,
            linkcolor=blue,
            anchorcolor=blue,
            citecolor=blue
            ]{hyperref}
\usepackage{doi}

\begin{document}
%\begin{CJK*}{UTF8}{song}
%\begin{CJK}{GBK}{song}

\title{Effects of tensor forces in nuclear spin-orbit splittings from ab initio calculations}% Force line breaks with \\

\author{Shihang Shen}
\affiliation{State Key Laboratory of Nuclear Physics and Technology, School of Physics,
Peking University, Beijing 100871, China}
\affiliation{RIKEN Nishina Center, Wako 351-0198, Japan}

\author{Haozhao Liang}
 \affiliation{RIKEN Nishina Center, Wako 351-0198, Japan}
 \affiliation{Department of Physics, Graduate School of Science, The University of Tokyo, Tokyo 113-0033, Japan}
\author{Jie Meng\footnote{Email: mengj@pku.edu.cn}}
 \affiliation{State Key Laboratory of Nuclear Physics and Technology, School of Physics,
Peking University, Beijing 100871, China}
 \affiliation{Yukawa Institute for Theoretical Physics, Kyoto University, Kyoto 606-8502, Japan}
 \affiliation{Department of Physics, University of Stellenbosch, Stellenbosch, South Africa}
\author{Peter Ring}
 \affiliation{State Key Laboratory of Nuclear Physics and Technology, School of Physics,
Peking University, Beijing 100871, China}
 \affiliation{Physik-Department der Technischen Universit\"at M\"unchen, D-85748 Garching, Germany}
\author{Shuangquan Zhang}
 \affiliation{State Key Laboratory of Nuclear Physics and Technology, School of Physics,
Peking University, Beijing 100871, China}

\date{\today}

\begin{abstract}
A systematic and specific pattern due to the effects of the tensor forces is found in the evolution of spin-orbit splittings in neutron drops.
This result is obtained from relativistic Brueckner-Hartree-Fock theory using the bare nucleon-nucleon interaction. It forms an important guide for future microscopic derivations of relativistic and nonrelativistic nuclear energy density functionals.
\end{abstract}


\maketitle

%==================Introduction======================================

The understanding of nuclear density functionals in terms of the nucleon-nucleon ($NN$) interaction is one of the present frontiers in nuclear physics. As
manifested by the quadrupole moment of the deuteron \cite{Bethe1940_PRC57-390}, the tensor force is an important component in the $NN$ interaction.
In the form of the two pion exchange the tensor force also provides the main part of the nuclear attraction \cite{Brown2010},
which is taken into account by the scalar $\sigma$ meson in phenomenological models \cite{Walecka1974}.
However, the role of the tensor force on the spin properties in finite nuclei is much less clear.

In configuration interaction (CI) calculations it has been found that the tensor force plays an important role in the shell structure far away from stability~\cite{Otsuka2005_PRL95-232502}. On the other side, in nearly all of the successful applications of phenomenological nuclear energy density functionals~\cite{Bender2003_RMP75-121}, tensor forces have been neglected for many years.

This has changed recently and much work has been done to investigate the impact of tensor forces in phenomenological nonrelativistic~\cite{Stancu1977_PLB68-108,Brown2006_PRC74-061303,Colo2007_PLB646-227,Brink2007_PRC75-64311,
Lesinski2007_PRC76-014312,ZOU-W2008-PRC77-014314,Zalewski2008_PRC77-024316,Moreno-Torres2010,Hellemans2012_PRC85-14326,
RocaMaza2013_PST154-014011,Grasso2014_PRC89-034316,Yuksel2014_PRC89-064322,Sagawa2014_PPNP76-76,Otsuka2006_PRL97-162501,Anguiano2011_PRC83-064306,
Anguiano2012_PRC86-054302,Grasso2013_PRC88-054328,Anguiano2016_EPJA52-183,Nakada2013_PRC87-067305}, and relativistic density functionals~\cite{Long2006_PLB640-150,Long2007_PRC76-034314,Long2008_EPL82-12001,Tarpanov2008,Lalazissis2009_PRC80-041301,
Marcos2014_PAN77-299,JIANG-LJ2014_PRC91-034326,LI-JJ2016_PLB753-97,Karakatsanis2017_PRC95-034318}.
Still, it is difficult to find significant features in experimental data which are only connected to tensor forces and therefore suitable for an adjustment of their parameters.
In a fit to nuclear masses and radii, for example, with relativistic density functional theory~\cite{Lalazissis2009_PRC80-041301}, one obtains the best fit for vanishing tensor forces.
On the other hand it has been found, that the single particle energies~\cite{Otsuka2005_PRL95-232502,Brown2006_PRC74-061303,Grasso2015_PRC92-054316} depend in a sensitive way on tensor forces.
However, in the context of density functional theory, single particle energies are only defined as auxiliary quantities~\cite{Kohn1965_PR140-A1133}. In experiment they are often fragmented and therefore only indirectly accessible. The fragmentation is caused by effects going beyond mean field, i.e., by the admixture of complicated configurations, such as the coupling to low-lying surface vibrations~\cite{Hamamoto1970_NPA141-1,Ring1973_NPA211-198,
Bortignon1977_PR30-305,Litvinova2006_PRC73-044328,Colo2010_PRC82-064307,Afanasjev2015_PRC92-044317}.

Obviously, the attempts to determine precise values for the strength parameters of the tensor forces in universal nuclear energy density functionals by a phenomenological fit to experimental data in finite nuclei is still a difficult problem~\cite{RocaMaza2013_PST154-014011}.
In such a situation we propose to determine these strength parameters from microscopic \emph{ab initio} calculations based on the well known bare nucleon-nucleon forces. In fact, much progress has been achieved in the microscopic description of nuclear structure in recent years~\cite{Dickhoff2004_PPNP52-377,LEE-D2009_PPNP63-117,LIU-Lang2012_PRC86-014302,
Barrett2013_PPNP69-131,Hagen2014_RPP77-096302,Carlson2015_RMP87-1067, Hergert2016_PR621-165,SHEN-SH2016_ChPL33-102103,SHEN-SH2017_PRC96-014316}. However, these are calculations of extreme numerical complexity and therefore they could be applied, so far, only in the region of light nuclei or for nuclei close to magic configurations.

For the investigation of heavy nuclei all over the periodic table, one is still bound to various versions of phenomenological nuclear density functionals and their extensions beyond mean field~\cite{Ring2009_PAN72-1285,Niksic2011_PPNP66-519,Meng2016}.
Of course the ultimate goal is an \emph{ab initio} derivation of such functionals. At present, such attempts are in their infancy~\cite{Negele1972_PRC5-1472,Drut2010_PPNP64-120,Dobaczewski2016}.
In Coulombic systems, where there exist very successful microscopically derived density functionals, one starts from the infinite system and the exact solution of an electron gas~\cite{Perdew2004_LNP620-269}.
In nuclei, there are attempts to proceed in a similar way and to derive in a first step semi-microscopic functionals. Modern relativistic and nonrelativistic \emph{ab initio} descriptions of symmetric nuclear matter at various densities are used as meta-data in order to reduce the number of phenomenological parameters of the density functionals considerably~\cite{Fayans1998_JETPL68-169,Baldo2010_JPG37-064015,RocaMaza2011_PRC84-054309}.
However, microscopic calculations of nuclear matter give us no information about the effective tensor force in the nuclear medium, because this is a spin-saturated system and their influence is therefore negligible. In order to learn the tensor force we propose in this letter to start from meta-data for a finite system, neutron drops confined in an external potential to keep the neutrons bound.

A neutron drop provides an ideal and simple system to investigate the neutron-rich environment.
Because of the missing proton-neutron interaction, the equations for neutron drops are much easier to be solved than those for finite nuclei. Therefore they have been investigated in the literature by many different \emph{ab initio} methods~\cite{Pudliner1996_PRL76-2416,Smerzi1997_PRC56-2549,Pederiva2004_NPA742-255,
Gandolfi2011_PRL106-012501,Bogner2011_PRC84-044306,Maris2013_PRC87-054318,
Potter2014_PLB739-445,Tews2016_PRC93-024305} and also by phenomenological density functional theory~\cite{ZHAO-PW2016_PRC94-041302}.

Starting from bare nuclear forces, Brueckner-Hartree-Fock (BHF) theory provides the $G$-matrix, a density dependent effective interaction in the nuclear medium and the basis of phenomenological density functional theory in nuclei~\cite{Vautherin1972_PRC5-626}.
As an \emph{ab initio} theory, the nonrelativistic version of BHF with 2N forces failed~\cite{Negele1970_PRC1-1260} because of the missing 3N forces, but it has been shown that the relativistic version allows to derive the saturation properties of infinite nuclear matter from bare 2N forces only~\cite{Anastasio1980_PRL45-2096}.

In this Letter we use the relativistic Brueckner-Hartree-Fock (RBHF) theory to study the effects of tensor forces in neutron drops.
This theory has recently been developed to describe finite nuclei self-consistently~\cite{SHEN-SH2016_ChPL33-102103,SHEN-SH2017_PRC96-014316} with results in much better agreement with experimental data than the nonrelativistic calculations based on 2N forces only.

We start from the relativistic bare nucleon-nucleon interaction Bonn A~\cite{Machleidt1989_ANP19-189} and investigate neutron drops confined in an external harmonic oscillator potential using the RBHF theory. We study drops with an even number of neutrons from $N = 4$ to $50$ and compare their energies and radii with other nonrelativistic \emph{ab initio} calculations.
Special attention is paid on the possible signature of the tensor force in the neutron-neutron interaction, that is, the evolution of spin-orbit (SO) splitting with neutron number.

%================== Theoretical Framework =============================

We start with a relativistic one-boson-exchange $NN$ interaction which describes the $NN$ scattering data~\cite{Machleidt1989_ANP19-189}. The Hamiltonian can be expressed as:
\begin{equation}
H=\sum_{kk^{\prime }}\langle k|T|k^{\prime }\rangle b_{k}^{\dagger
}b_{k^{\prime }}^{{}}+\frac{1}{2}\sum_{klk^{\prime }l^{\prime }}\langle
kl|V|k^{\prime }l^{\prime }\rangle b_{k}^{\dagger }b_{l}^{\dagger
}b_{l^{\prime }}^{{}}b_{k^{\prime }}^{{}},  \label{eq:hami}
\end{equation}%
where the relativistic matrix elements are given by
\begin{align}
\langle k|T|k^{\prime }\rangle & =\int d^{3}r\,\bar{\psi}_{k}(\mathbf{r})\left( -i%
\bm{\gamma}\cdot \nabla +M\right) \psi _{k^{\prime }}(\mathbf{r}), \\
\langle kl|V_{\alpha }|k^{\prime }l^{\prime }\rangle & =\int
d^{3}r_{1}d^{3}r_{2}\,\bar{\psi}_{k}(\mathbf{r}_{1})\Gamma _{\alpha }^{(1)}\psi
_{k^{\prime }}(\mathbf{r}_{1}) \label{eq:V} \notag \\
&~~~~~~~~\times D_{\alpha }(\mathbf{r}_{1},\mathbf{r}_{2})\bar{\psi}_{l}(\mathbf{r}_{2})\Gamma
_{\alpha }^{(2)}\psi _{l^{\prime }}(\mathbf{r}_{2}).
\end{align}
The indices $k,l$ run over a complete basis of Dirac spinors with positive and negative energies, as, for instance, over the eigensolutions of a Dirac equation with potentials of Woods-Saxon shape~\cite{ZHOU-SG2003_PRC68-034323,SHEN-SH2017_PRC96-014316}.

The two-body interaction $V_\alpha$ contains the exchange contributions of different mesons $\alpha= \sigma,\delta, \omega,\rho, \eta, \pi$. The interaction vertices $\Gamma _{\alpha }$ for particles 1 and 2 contain the corresponding $\gamma$-matrices for scalar $(\sigma,\delta)$, vector $(\omega,\rho)$, and pseudovector $(\eta, \pi)$ coupling and the isospin matrices $\vec\tau$ for the isovector mesons $\delta, \rho,$ and $\pi$.
For the Bonn interaction \cite{Machleidt1989_ANP19-189}, a form factor of monopole-type is attached to each vertex and $D_{\alpha }(\mathbf{r}_{1},\mathbf{r}_{2})$ represents the corresponding meson propagator. Retardation effects were deemed to be small and were ignored from the beginning.
Further details are found in Ref.~\cite{SHEN-SH2017_PRC96-014316}.

The matrix elements of the bare nucleon-nucleon interaction are very large and difficult to be used directly in nuclear many-body theory. Within Brueckner theory, the bare interaction is therefore replaced by an effective interaction in the nuclear medium, the $G$-matrix. It takes into account the short-range correlations by summing up all the ladder diagrams of the bare interaction~\cite{Brueckner1954_PR95-217,Brueckner1954_PR96-508} and it is deduced from the Bethe-Goldstone equation~\cite{Bethe1957_PRSA238-551},
\begin{equation}
\bar{G}_{aba'b'}(W) = \bar{V}_{aba'b'}
 +\frac{1}{2}\sum_{cd} \frac{\bar{V}_{abcd}\bar{G}_{cda'b'}(W)}{W-\varepsilon _{c}-\varepsilon _{d}},
\label{eq:BG}
\end{equation}%
where in the RBHF theory $|a\rangle,|b\rangle$ are solutions of the relativistic Hartree-Fock (RHF) equations, $\bar{V}_{aba'b'}$ are the anti-symmetrized two-body matrix elements (\ref{eq:V}) and $W$ is the starting energy. The intermediate states $c,\,d$ run over all states above the Fermi surface with $\varepsilon_c,\,\varepsilon_d > \varepsilon_F$.

The single-particle motion fulfills the RHF equation in the external field of a harmonic oscillator (HO):
\begin{equation}
(T+U+\frac{1}{2} M\omega^2 r^2)|a\rangle =e_{a}|a\rangle ,
\label{eq:rhf}
\end{equation}%
where $e_{a}=\varepsilon _{a}+M$ is the single-particle energy with the rest
mass of the nucleon $M$ and $\hbar\omega = 10$ MeV.
The self-consistent single-particle potential $U$ is defined  by the $G$-matrix \cite{Baranger1969_Varenna40,Davies1969_PRC177-1519,SHEN-SH2017_PRC96-014316}:
\begin{equation}
\langle a|U|b\rangle = \sum_{c=1}^{N}\langle ac|\bar{G}|bc\rangle ,
\label{eq:UG}
\end{equation}
where the index $c$ runs over the occupied states in the Fermi sea (\emph{no-sea} approximation). In contrast to the RBHF calculations for self-bound nuclei in Ref.~\cite{SHEN-SH2016_ChPL33-102103,SHEN-SH2017_PRC96-014316}, a center of mass correction is not necessary in the external field.

The coupled system of RBHF equations (\ref{eq:BG}), (\ref{eq:rhf}), and (\ref{eq:UG}) is solved by iteration. The initial basis is a Dirac Woods-Saxon basis \cite{ZHOU-SG2003_PRC68-034323} obtained by solving the spherical Dirac equation in a box with the size $R_{\rm box} = 8$ fm and a mesh size $dr=0.05$~fm. During the RBHF iteration it is gradually transformed to the self-consistent RHF basis as explained in Ref.~\cite{SHEN-SH2017_PRC96-014316}.
The Bethe-Goldstone equation~(\ref{eq:BG}) is solved in the same way as in Ref.~\cite{SHEN-SH2017_PRC96-014316}, except that now only the isospin channel $T_z = 1$ is included.

\begin{figure}
\includegraphics[width=8cm]{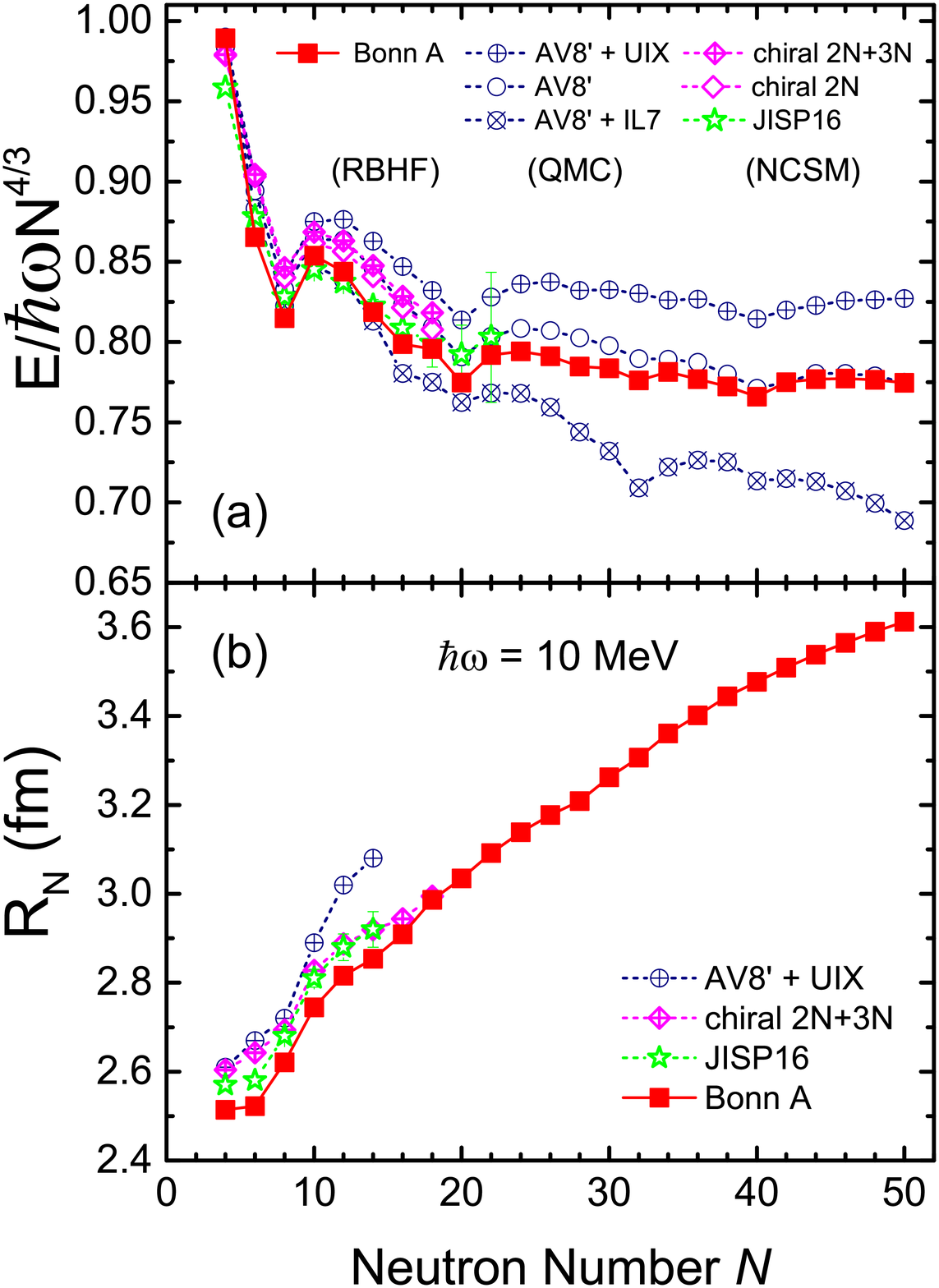}
\caption{(Color online) (a) Total energy divided by $\hbar\omega N^{4/3}$ and (b) radii of $N$-neutron drops in a HO trap calculated by the RBHF theory using the interaction Bonn A~\cite{Machleidt1989_ANP19-189},
in comparison with other nonrelativistic \emph{ab initio} calculations. See text for details.}
\label{fig1}
\end{figure}

Fig.~\ref{fig1} shows the total energy $E$ in units of $\hbar\omega N^{4/3}$ and the radii of $N$-neutron drops (with $N$ from 4 to 50) in a HO trap calculated by the RBHF theory using the bare interaction Bonn A~\cite{Machleidt1989_ANP19-189}. For the cases of  open shells, the filling approximation is used.

The results are compared with the quantum Monte-Carlo (QMC) calculations~\cite{Gandolfi2011_PRL106-012501,Maris2013_PRC87-054318} based on the 2N interaction AV8'~\cite{Pudliner1997_PRC56-1720} (without and with the 3N forces UIX and IL7), with the no-core shell model (NCSM) calculations~\cite{Potter2014_PLB739-445,Maris2013_PRC87-054318} based on the chiral 2N + 3N forces, and the force JISP16.
The factor $\hbar\omega N^{4/3}$ takes into consideration that in the Thomas-Fermi approximation~\cite{Ring1980} the total energy for a non-interacting $N$-Fermion system in a HO trap is given by
\begin{equation}\label{eq:Etf}
E = \frac{3^{4/3}}{4}\hbar\omega N^{4/3}\approx 1.082\,\hbar\omega N^{4/3}.
\end{equation}
With increasing neutron number of the drops we observe a saturation of $E/\hbar\omega N^{4/3}$ for $N\geq 20$, in contrast to the nuclear case where the binding energy per nucleon saturates for large mass number $A$.

By comparing with the QMC and NCSM calculations in panel (a), the results of the RBHF with the interaction Bonn A are similar to those obtained with the JISP16 interaction. For $N \leq 14$, Bonn A is also similar to AV8' + IL7, but getting closer to AV8' afterwards.
This result is favourable as JISP16 is a phenomenological nonlocal $NN$ interaction which reproduces the scattering data as well as it gives a good description for light nuclei \cite{Shirokov2007_PLB644-33,Maris2009_PRC79-014308}.
On the other hand, AV8' + IL7 gives a much better description for light nuclei up to $A = 12$ than AV8' or AV8' + UIX, but the three-pion rings included in IL7 give too much over-binding for pure neutron matter at higher densities \cite{Sarsa2003_PRC68-024308,Maris2013_PRC87-054318}.

Panel (b) of Fig.~\ref{fig1} shows the corresponding radii.
While the energies of RBHF with Bonn A are similar to those of the JISP16 interaction, the radii of RBHF are smaller. In comparison with the results of AV8' + UIX and chiral force, the energies and radii of RBHF with Bonn A are smaller, except when $N$ approaches 18, where the radii become close to chiral 2N + 3N results.

\begin{figure}
\includegraphics[width=8cm]{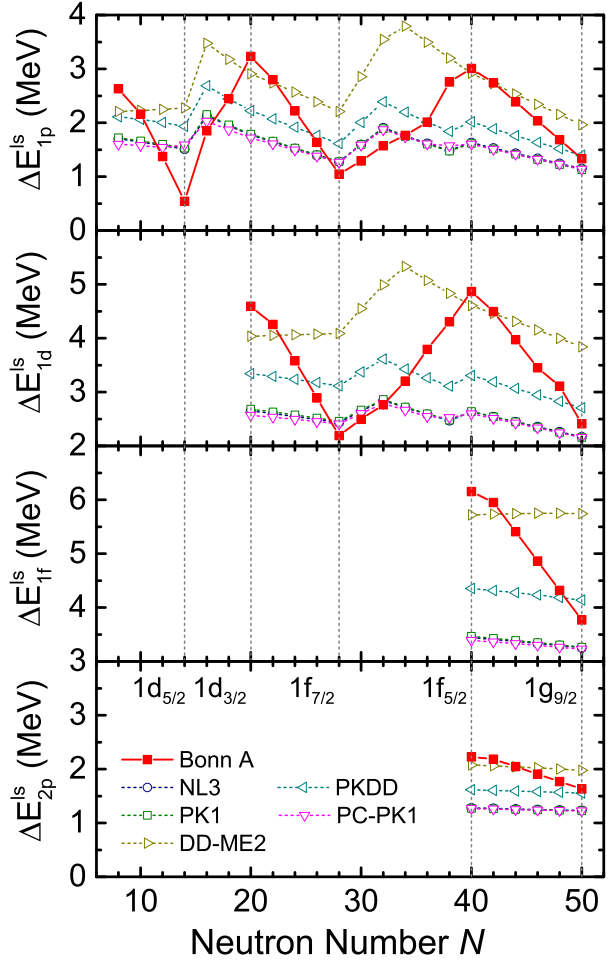}
\caption{(Color online) From top to bottom panel, $1p,\,1d,\,1f$, and $2p$ spin-orbit splittings of $N$-neutron drops in a HO trap ($\hbar\omega = 10$ MeV) calculated by the RBHF theory using the Bonn A interaction, in comparison with the results obtained by various RMF density functionals.}
\label{fig2}
\end{figure}

In Fig.~\ref{fig2}, we show the SO splittings of $N$-neutron drops for $1p,\,1d,\,1f$, and $2p$ in a HO trap calculated by the RBHF theory using the Bonn A interaction.
They are compared with results obtained by  various phenomenological relativistic mean-field (RMF) density functionals, including the nonlinear meson-exchange models NL3~\cite{Lalazissis1997_PRC55-540} and PK1~\cite{Long2004_PRC69-034319}, the density-dependent meson-exchange models DD-ME2~\cite{Lalazissis2005_PRC71-024312} and PKDD \cite{Long2004_PRC69-034319}, and the nonlinear point-coupling model PC-PK1~\cite{ZHAO-PW2010_PRC82-054319}.
This figure shows the evolution of the various SO splittings with neutron number.
For the microscopic RBHF results we find a clear pattern: The SO splitting of a specific orbit with orbital angular momentum $l$ decreases as the next higher $j=j_>= l + 1/2$ orbit is filled and reaches a minimum when this orbit is fully occupied. As the number of neutron continues to increase, the $j=j_< = l - 1/2$ orbit begins to be occupied and the SO splitting increases.

Otsuka et al.~\cite{Otsuka2005_PRL95-232502} have found a similar effect between neutron and proton in nuclei. They explained it in terms of the monopole effect of the tensor force, which produces an attractive interaction between a proton in a SO aligned orbit with $j=j_>=l + 1/2$ and a neutron in a SO anti-aligned orbit with $j'=j'_<=l' - 1/2$ and a repulsive interaction between the same proton and a neutron in a SO aligned orbit with $j'=j'_>=l'+ 1/2$.

\begin{figure}[!thbp]
\includegraphics[width=8cm]{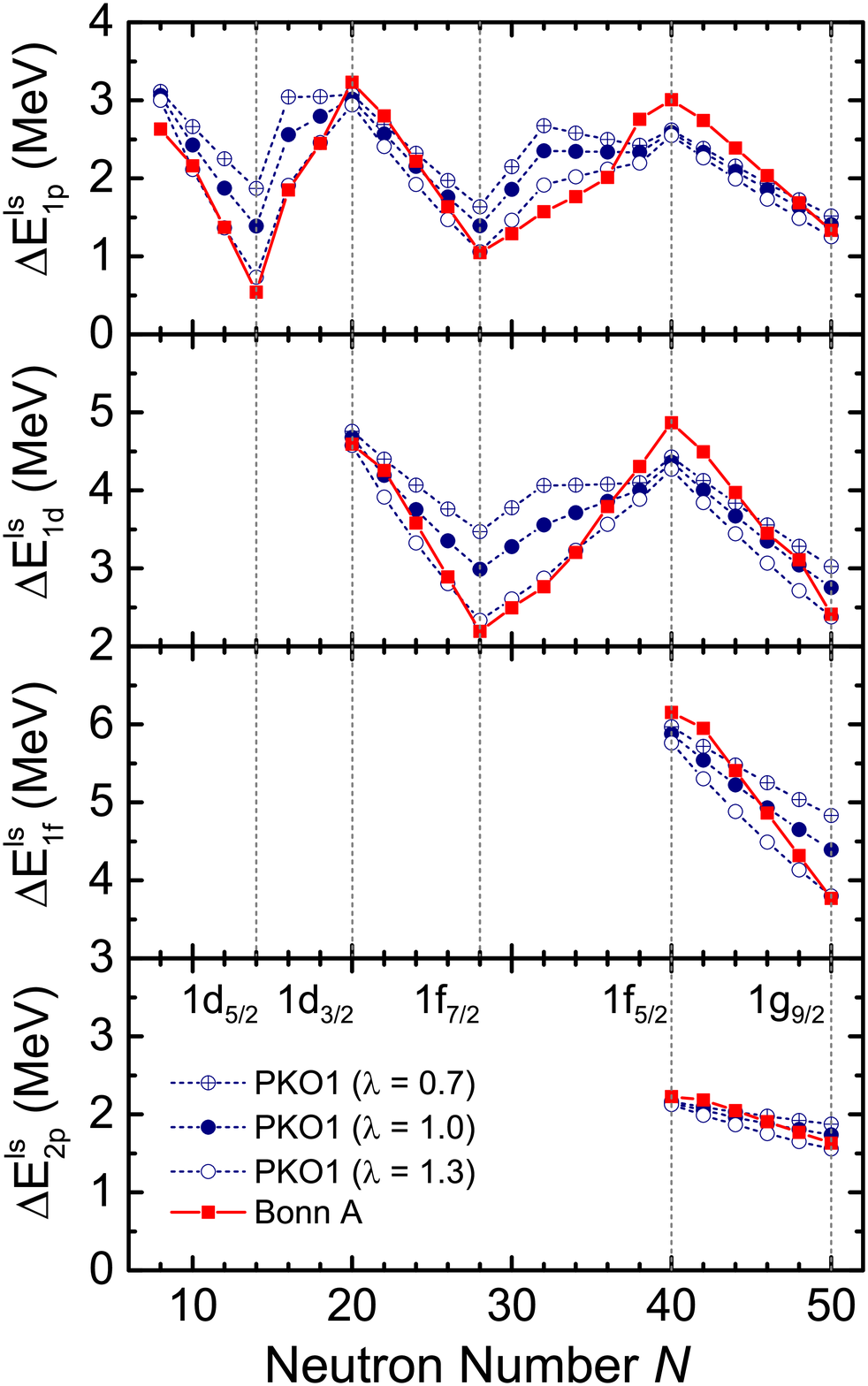}
\caption{(Color online) Similar to Fig.~\ref{fig2}, but in comparison with the RHF density functional PKO1 \cite{Long2006_PLB640-150} with different strength of pion coupling characterized by $\lambda$.}
\label{fig3}
\end{figure}

As discussed in Ref.~\cite{Otsuka2005_PRL95-232502} a similar mechanism, but with smaller amplitude, exists also for the tensor interaction between neutrons with $T=1$.
Therefore we can explain the behavior of the SO splitting in Fig.~\ref{fig2} in a qualitative way: we consider, for instance, the decrease of the $1d$ SO splitting if we go from $N=20$ to $N=28$.
 Because of the interaction with the neutrons filling the $1f_{7/2}$ shell above $N=20$, the $1d_{5/2}$ orbit is shifted upward and the  $1d_{3/2}$ is shifted downward, thus reducing the $1d$ SO splitting.
  Above $N=28$ we fill in neutrons into $2p_{1/2}$ and $1f_{5/2}$. They interact with the $1d$-neutrons in the opposite way and increase the SO-splitting for the $1d$ configuration.

On the other hand, this specific evolution of SO splitting is not significant for any of the
phenomenological RMF density functionals in Fig.~\ref{fig2}, which do not include a tensor term.
In order to verify that this specific pattern is indeed caused by the tensor term, we show in Fig.~\ref{fig3} the same calculation but with the RHF density functional PKO1 \cite{Long2006_PLB640-150}, which includes the tensor force induced by the pion coupling through the exchange term.
Without readjusting the other parameters of this functional, we have multiplied a factor $\lambda$ in front of the pion coupling to investigate the effects of the tensor forces.
It is remarkable to see that the evolution of the SO splitting is influenced by
the strengths of the tensor forces significantly. For $\lambda=1$ we have the results of the density
functional PKO1. They show already the right pattern, but the size of the effect is somewhat too small. This can be understood by the fact, that it is difficult to fit the strengths of the tensor forces just to bulk properties such as binding energies and radii~\cite{Lalazissis2009_PRC80-041301}.
The general feature of these SO splittings found in our RBHF calculations with Bonn A can be well reproduced with PKO1 simply by multiplying a factor $\lambda = 1.3$ in front of the pion coupling.
One may wonder if the tensor force discussed here is too strong, as the original functional PKO1 can well reproduce the experimental data of SO splitting reduction in realistic nuclei, e.g., from $^{40}$Ca to $^{48}$Ca \cite{Long2007_PRC76-034314}.
However, as pointed out in the introduction, it is never easy to compare directly to the experimental single-particle energies in realistic nuclei, as complicated beyond-mean-field effects are involved.

Of course, finally one should carry out a complete fit, taking into account at the same time the \emph{ab-initio} meta data for nuclear matter as well as these meta-data for SO splittings in neutron drops together with a fine-tuning of a few final parameters to masses and radii. Work in this direction is in progress, but it goes definitely beyond the scope of this letter.

Finally we would like to mention two important aspects of these results:

(a) The specific pattern of increasing and decreasing SO splittings with neutron number is not restricted to a specific $j$-shell, i.e., to a specific region. It seems to be generally valid for all the neutron numbers $4 \leq N \leq 50$ under investigation and it can be reproduced by readjusting a single parameter $\lambda$ for the tensor strength in the density functional PKO1. Therefore we can expect that a similar feature is valid also for real nuclei all over the periodic table.

(b) In relativistic Brueckner-Hartree-Fock theory, there are no higher-order configurations~\cite{Day1967_RMP39-719}.
This means, that effects like particle-vibrational coupling are not included in these meta-data.
This allows to adjust the tensor force to these meta-data without the ambiguity of additional effects of particle vibrational coupling. They are neither included in
the present concept of density functional theory, nor in the present RBHF calculations. In the case of realistic calculations in finite nuclei comparable with experimental data, they have to be included by going beyond mean field in a similar way as in Ref. \cite{Litvinova2006_PRC73-044328}

In summary, we have studied neutron drops confined in an external field of oscillator shape using the relativistic Brueckner-Hartree-Fock theory with the bare $NN$ interaction.
It was found that the SO splitting decreases as the next $j = l + 1/2$ orbit being occupied, and increases again as the next $j = l - 1/2$ orbit being occupied.
This is similar to the effects of tensor forces between neutron and proton as has been found in Ref.~\cite{Otsuka2005_PRL95-232502}.
The pattern of the evolution of SO splittings cannot be reproduced by the RMF density functionals, while it can be well reproduced with the RHF density functional PKO1 which includes tensor forces. This implies that the strengths of tensor forces in neutron drops can be derived from \emph{ab initio} calculations and used as a guide for future \emph{ab initio} derivations of nuclear density functionals.

\section*{ACKNOWLEDGMENTS}

We thank Pengwei Zhao for discussions and providing his results.
This work was partly supported by the Major State 973 Program of China No.~2013CB834400, Natural Science Foundation of China under Grants No.~11335002, No.~11375015, and No.~11621131001,  the Overseas Distinguished Professor Project from Ministry of Education of China No.
MS2010BJDX001, the Research Fund for the Doctoral Program of Higher Education of China under Grant No.~20110001110087, and the DFG (Germany) cluster of excellence \textquotedblleft Origin and Structure of the Universe\textquotedblright\ (www.universe-cluster.de). S.S. would like to thank the short-term Ph.D. student exchange program of Peking University and the RIKEN IPA project, and H.L. would like to thank the RIKEN iTHES project and iTHEMS program.

\bibliographystyle{apsrev4-1}
%\bibliography{n-drops,liter}

%\end{CJK}
%\end{CJK*}
\end{document}